# Time-resolved photoluminescence of n-doped SrTiO$_3$

A. Rubano, D. Paparo, M. Radović, A. Sambri, F. Miletto Granozio, U. Scotti di Uccio[a],
L. Marrucci[b].

*CNR-INFM Coherentia and Dipartimento di Scienze Fisiche, Università di Napoli "Federico II", Complesso di Monte S. Angelo, via Cintia, 80126 Napoli, Italy.*

[a] Also at DiMSAT, Università di Cassino, via Di Biase 43, 03043 Cassino (FR), Italy.
[b] Electronic mail: lorenzo.marrucci@na.infn.it

**Abstract:** Following the recent surge of interest in n-doped strontium titanate as a possible blue light emitter, a time-resolved photoluminescence analysis was performed on nominally pure, Nb-doped and oxygen-deficient single-crystal SrTiO$_3$ samples. The doping-effects on both the electronic states involved in the transition and the decay mechanism are respectively analyzed by comparing the spectral and dynamic features and the yields of the emission. Our time-resolved analysis, besides shedding some light on the basic recombination mechanisms acting in these materials, sets the intrinsic bandwidth limit of the proposed blue light emitting optoelectronic devices made of Ti-based perovskites heterostructures in the GHz range.

**PACS:** 78.55.-m ; 78.47.+p ; 71.35.Aa ; 71.35.Ee .

Strontium titanate, or SrTiO$_3$ (STO), plays a crucial role in the wide class of functional oxide materials with perovskite or perovskite-like structure, and it is often considered as the "prototypical" compound of this class. A broad range effort for understanding its multifunctional properties has been undertaken since a few decades by the community working on transition metal oxides. Beside the intrinsic interest for basic physics and possible applications, the comprehension of the properties of STO is often envisaged as a preliminary step to understand more complex compounds, with related crystal structure and strongly correlated electronic behaviour, as those exhibiting high T$_c$ superconductivity, magnetic and magnetoresistive behaviour, ferroelectric/multiferroic properties, and so on.

As a matter of fact STO has been found, as a "case study", to be itself everything but "simple". This perovskite shows, in fact, several unexpected and sometimes ill understood properties, such as quantum paraelectricity, record-high permittivity (up to $10^4$ at 10K) [1], ferroelectricity induced by compressive biaxial strain [2] or by isotopic substitution [3], ultrahigh carrier mobility in epitaxial heterostructures with LaAlO$_3$ [4] and very variable and controversial surface properties [5]. Furthermore, under the introduction of n-type dopants, including, in a generalized sense, oxygen vacancies ($V_O^{\bullet\bullet}$, in the Kröger-Vink notation), STO readily turns from a transparent insulator with a 3.2 eV indirect gap into an opaque, black/dark-blue material, with metallic [6] and superconducting [7] transport properties and a peculiar inhomogeneous behaviour [8]. The possibility of p-doping has also been demonstrated [9].

A totally new field of interest and application has emerged from recent papers [10,11] reporting a previously unnoticed fascinating optical feature of n-doped STO: a room-temperature broad-band blue light emission occurring under exposure to UV light (3.8 eV). In those works, the authors doped crystalline STO with electrons, by substitution of Nb$^{4+}$ for Ti$^{3+}$, La$^{3+}$ for Sr$^{2+}$, or introducing $V_O^{\bullet\bullet}$ sites by Ar$^+$-ion irradiation. According to the authors, all such treatments turned stoichiometric samples lacking any emitting properties in the visible domain (at room temperature) into blue-light emitters, all essentially showing the same spectral features [11]. It was shown that the different emitting behaviour of doped and pure STO allowed to pattern the blue-light emitting region into any size and shape, by combining conventional photolitography and Ar$^+$-milling technique [10].

In view of the renew of interest raised by these findings, and in order to determine their real potential, two questions need to be tackled: (i) How far are the reported blue emitting properties of n-doped STO actually related to the presence of impurities or vacancies acting as donors? In fact, while refs [10] and [11] seem to provide conclusive evidence that it is indeed the "oxygen vacancies shining blue" [12], according to other reports blue light emission is also found in intrinsic STO [13]. This was recently confirmed by some of the authors [14]. (ii) How suitable is the discovered luminescence for high frequency applications, and what is the intrinsic limit to the bandwidth of the potential future devices?

In order to investigate these issues, in this work time-resolved photoluminescence (PL) measurements were performed on "intrinsic" ("*I*"), Nb-doped ("*Nb*") and oxygen deficient ("*O*") STO samples. With regard to the first quoted issue, time resolution complements spectral resolution in allowing the comparison of the mechanisms taking place in different samples. While the spectral analysis allows us to compare the energy of the electronic states involved in the luminescence, the analysis of the temporal profiles allows us to compare the mechanisms (both radiative and non radiative) causing the decay of the excited plasma. With regard to the second issue, the intrinsic bandwidth limit can be trivially estimated to be of the order of the inverse of the decay half-life.

The samples employed in this work were commercial (100) oriented single crystals at room temperature. *I* samples where pure, as-received, STO; *O* samples were obtained from *I* samples by annealing them for 1 h at 950 °C and $10^{-9}$ mbar (base pressure $10^{-11}$ mbar); *Nb* samples were commercial doped crystals with a 0.2% mol Nb concentration. Every measurement was repeated on more samples. The procedure adopted for *O* samples is known to introduce $V_O^{\bullet\bullet}$ sites and changed the transparent insulating *I* samples into black/dark-blue, opaque, conducting samples. In order to estimate the doping-induced carrier density of these *O* samples, we measured the room temperature resistivity of a 15 nm thick STO film grown on $NdGaO_3$ and annealed in the same conditions. We found a resistivity of 3.4 Ω·cm, corresponding to a carrier density of about $3\times10^{17}$ cm$^{-3}$ (see Fig. 1 in Ref. [15]). The Nb→Ti substitution is known to induce a local energy level within the band-gap very close to the conduction band [16], while the lack of oxygen induces $Ti^{3+}$-$V_O^{\bullet\bullet}$ complexes [17] in which the $Ti^{3+}$ ions are deep-hole-traps and the $V_O^{\bullet\bullet}$ sites are probably shallow-electron-traps.

The experimental setup is described in ref [14]. Briefly, for excitation we used 25-ps UV pulses at a wavelength of 355 nm (3.49 eV), with a fluence ranging up to 45 mJ/cm$^2$, corresponding to an estimated photogenerated electron-hole pair density of $6\times10^{20}$ cm$^{-3}$ within an optical penetration length. Time-resolved emission was detected employing a 5 GHz bandwidth photodiode+digital oscilloscope system. In our experiments, no damage on the sample surfaces was ever seen at the end of every measurement run, and the behaviour of each given sample was found to be highly reproducible and reversible. This makes us confident that light exposure does not alter significantly the oxygen-vacancy concentration.

In the inset of Fig. 1, the normalized luminescence spectra of *I*, *Nb* and *O* samples measured in the same experimental conditions (same geometry, beam energy and focusing) are shown. Spectra are vertically shifted for clarity. It is seen that the responses of the differently doped samples, at least as far as radiative decay processes are concerned, are largely overlapping. The fluorescence yields measured with the three different kinds of samples are comparable, and their differences are anyway grossly within the sample-to-sample variations occurring among *I* samples from different vendors. All spectra are peaked at about 425 nm (photon energy of 2.9 eV) and are comparable to those reported in refs [10,11], although a relative depletion of the intensity is found in our data in the spectral range between 450 and 500 nm. Our measurements taken as a function of pump pulse energy (not shown) show that the spectrum shape is intensity-independent, and that its integral scales linearly with pulse energy, with no sign of saturation for energies ranging up to 1 mJ per pulse.

In the main panel of Fig. 1, three examples of decay curves collected from *I*, *Nb* and *O* samples are reported, for comparison, in log scale. The overall dynamics is found to be very similar in different samples, although the decay appears to be faster in the *Nb* sample. The oscillations and other minor features are due to the instrumental response function [14]. The decay dynamics of the *I* samples has been addressed by some of the authors in [14], where it was shown to be consistent with the presence of two distinct relaxation processes taking place in the electron-hole plasma generated by the laser pulse (see Fig. 2). These two processes exhibit different rates and different decay laws. The faster process is a "bimolecular" decay (BD), with a relaxation rate and the corresponding luminescence intensity which are both proportional to the square of the electron-hole pair density. The slower process is a "unimolecular" decay (UD), with a rate and emission intensity which are linear in the density. We refer to ref [14] for all details on this model and on the fitting procedure. In this work, we applied the same UD+BD model (with the same "global" fitting procedure [14]) to the *Nb* and *O* doped samples, in order to determine the possible doping dependence of the underlying decay mechanism. An example of the fit results is shown in the inset of Fig 3. It was found that our model yields a very good fit to the decay dynamics, also as a function of the pulse energy, on all the considered samples. This provides strong evidence that the presence of doping does not qualitatively affect the basic physics of the decay mechanism. Some quantitative differences in the best-fit values of the model parameters were however found. Indeed, after taking into account the sample-to-sample variability seen within the same class of samples, we concluded that the *Nb* samples are characterised by an overall faster dynamics of both the UD and the BD term, while minor differences are found between the *I* and the *O* samples. The effect of Nb in accelerating the decay seems to be stronger for the UD term (by a factor 2-5), while the decay time of the BD term seems to be more weekly dependent on doping (with a typical factor of about 1.4). The different behaviour of *Nb* and *O* samples suggests that the role of Nb in accelerating the radiative decay processes might be independent of its role as a donor. The UD and BD fluorescence yields measured in doped samples are also comparable to intrinsic ones, as can be easily seen in Fig 3, where the UD and BD contributions to the decay integral are plotted as a function of the pulse energy. Note that the doped-sample yields are slightly lower than the *I* sample one, but the optical absorption of the emitted light in opaque doped samples could account for this minor difference.

In order to estimate the bandwidth of a potential STO-based light emitting device, the reciprocal of the overall full width at half maximum of the radiative decay profile was plotted in Fig. 4 versus the beam energy for a *Nb* sample. It is seen that the bandwidth, being mostly determined by the faster BD process, scales linearly with energy, ranging up to about 2 GHz at 1 mJ. This linear behavior is in good agreement with the model predictions [14]. In the upper horizontal scale, the pulse energy is converted into a gross estimate of pair density, based on the effective volume in which the pulse energy is absorbed. This scale has the advantage of being directly applicable also to hypothetical electroluminescent devices.

In summary, very strong overall similarities are found in the emission spectrum, the yield and the decay dynamics of the photoluminescent response of pure, oxygen deficient and Nb-doped STO. This results point to a very minor role of donors and doping induced electrons both on the states involved in the transition, and at least on the most relevant decay mechanism, i.e., the BD. In spite of the solid self-consistency, and of the agreement with the blue luminescence reported on pure STO in [13], our results seem to be in disagreement with the results reported by Kan et al. [11,10], who found namely no blue light emission from intrinsic STO. We speculate that this might be due to one of the following two reasons: *a)* The higher photon excitation energy employed by Kan et al. (3.8 eV) with respect to our case. In fact, photoluminescence excitation (PLE) spectra in pure STO [13] show a deep depletion of the spectrum for excitation energies exceeding 3.7 eV; *b)* The different energy density. In our pulsed high fluence regime, the density of photoinduced e-h pairs, $\sim 10^{20}/cm^3$, is at least one order of magnitude larger than the density of doping-induced carriers. In the case of ref [11,10], instead, the density of photoinduced carriers is presumably much lower, and the presence of doping-induced carriers might therefore play a crucial role.

Our results set the upper bandwidth limit of STO-based emitting devices above 1 GHz, opening some interesting prospects on the fabrication of integrated optoelectronic devices based on titanates, where the growth of epitaxial heterostructures allows integrating emitting elements with tunable filters, optical switches, and ultrawide bandwidth modulators [18].

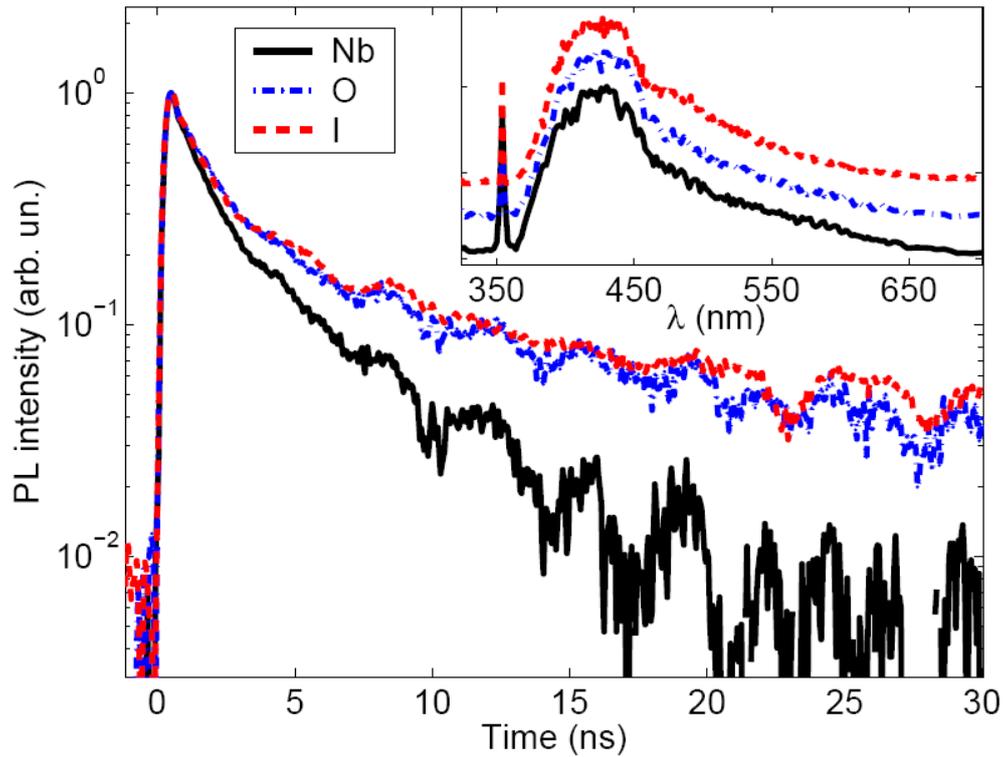

**Fig 1:** (Color online) Time-resolved decay curves of Nb-doped (*Nb*, black solid curve), oxygen deficient (*O*, blue dot-dashed curve) and intrinsic (*I*, red dashed curve) samples. The curves are normalized to their maximum and plotted in a semilogarithmic vertical scale. Oscillations and other minor features are due to the instrumental response function. The pulse energy is 0.5 mJ in all cases. Inset: photoluminescence spectra of the same samples. The spectra are normalized to their maximum and vertically shifted for clarity. The narrow peak seen at 355 nm is due to the residual UV scattering.

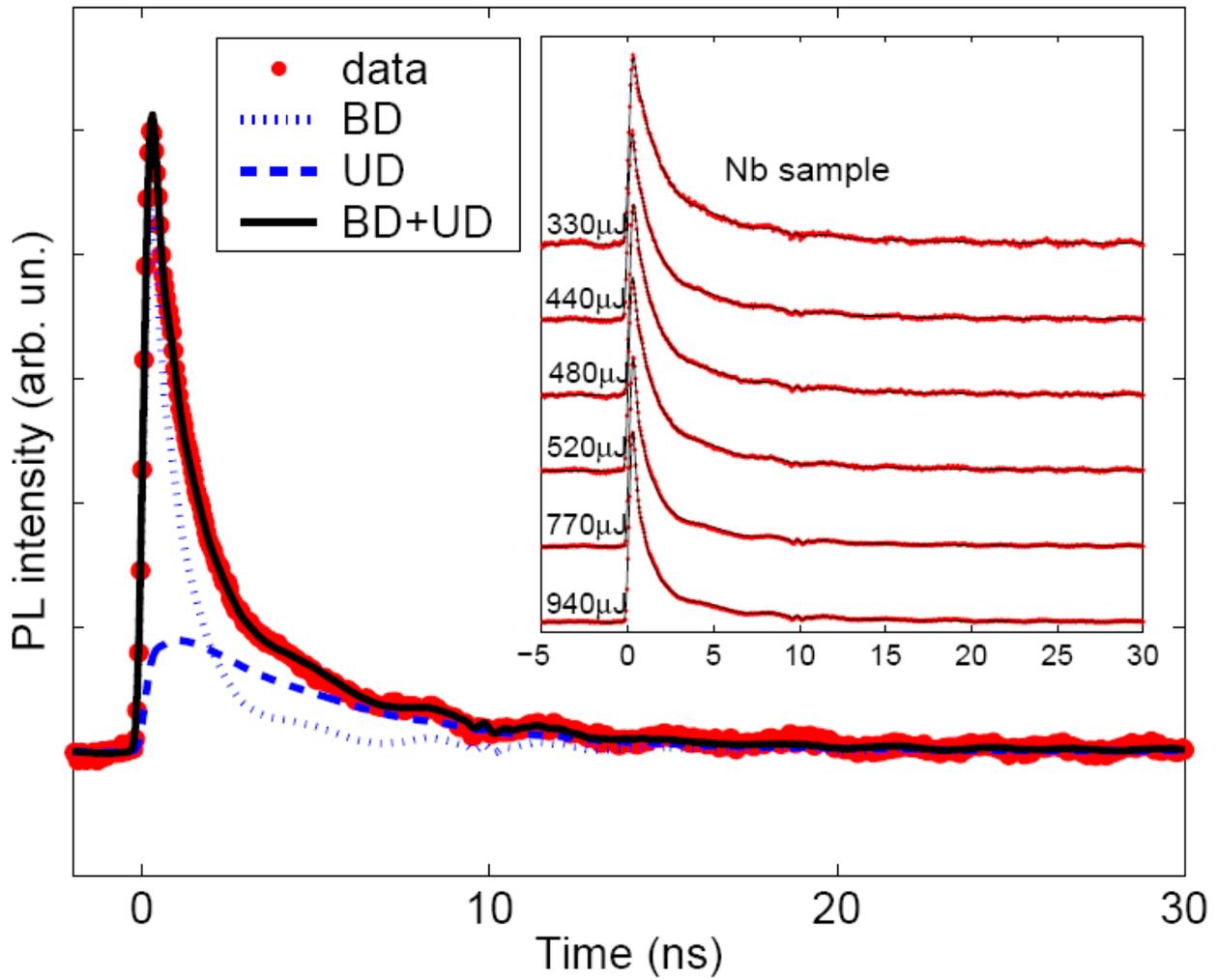

**Fig 2:** (Color online) An example of the global-fit curve in the case of a Nb sample. BD (blue dotted) and UD (blue dashed) decay components are plotted separately. Gray (red) circles are the data points. The black solid curve is the total fitting curve. Inset: the whole set of decays measured for different pump energies (red circles) and on which the global-fit was performed (black solid lines) is shown. The pulse energy corresponding to each curve is indicated on the left of each curve.

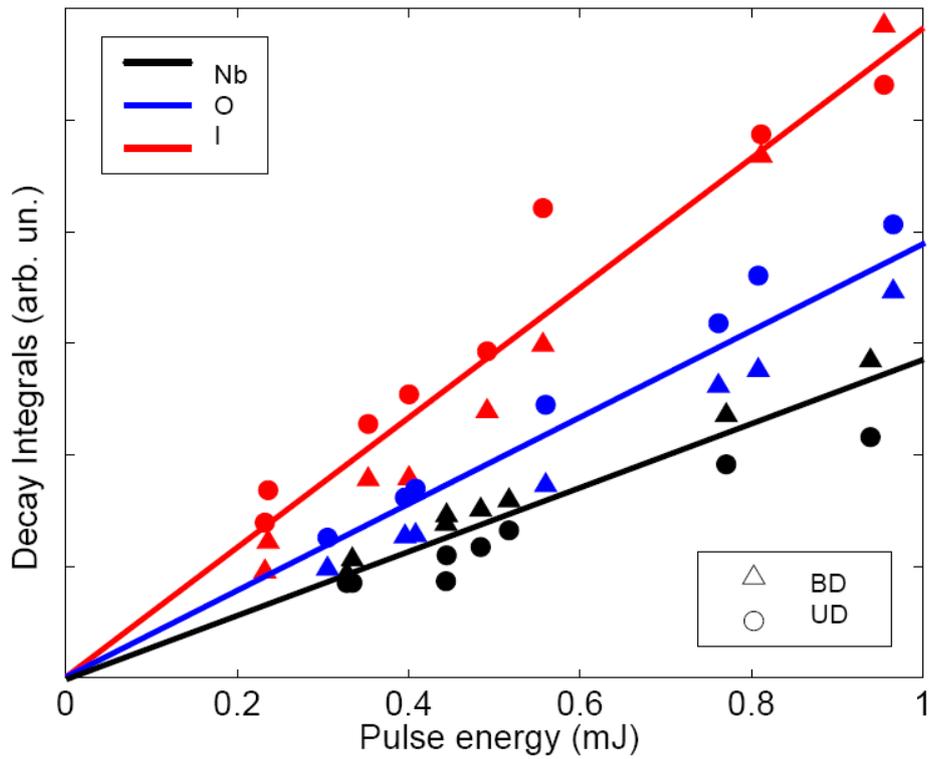

**Fig 3:** (Color online) Time-integrated decay luminescence intensity as a function of the pulse energy for *Nb* (black), *O* (dark-gray, blue online) and *I* (light gray, red online) samples. Circles are the UD decay integrals, triangles are the BD decay integrals. The slope of each line is the BD-UD yield. Note that all samples show approximately comparable yields for the UD and BD channels.

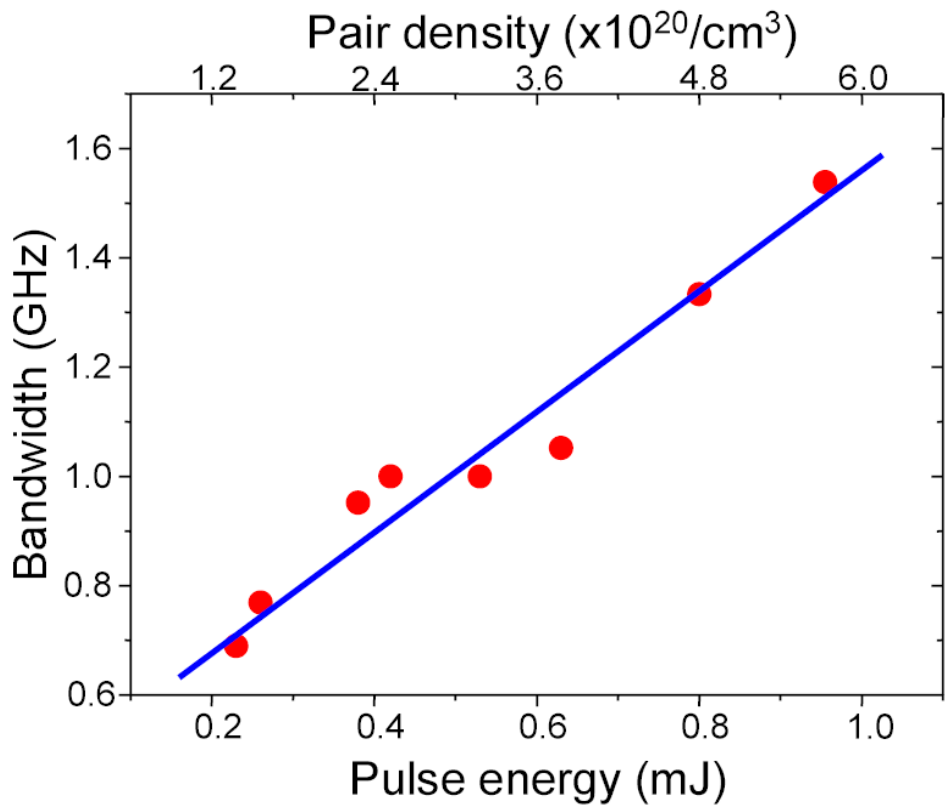

**Fig 4:** (Color online) The estimated bandwidth (reciprocal of the full width at half maximum) is found to be a linear function of the pulse energy and of the estimated pair density (upper scale). Gray (red online) circles are the data points, the (blue) solid line is the linear fit. The slope is about 1.06 GHz/mJ. The intercept (0.48 GHz) is not consistent with zero because at very low energies the decay is dominated by the UD lifetime, which is energy-independent.

**References**

[1] K.A. Muller, Phys. Rev. B **19**, 3593 (1979).

[2] J. H. Haeni, P. Irvin, W. Chang, R. Uecker, P. Reiche, Y. L. Li, S. Choudhury, W. Tian, M. E. Hawley, B. Craigo, A. K. Tagantsev, X. Q. Pan, S. K. Streiffer, L. Q. Chen, S. W. Kirchoefer, J. Levy, D. G. Schlom, Nature **430**, 758 ( 2004).

[3] M. Itoh, R. Wang, Y. Inaguma, T. Yamaguchi, Y-J. Shan, and T. Nakamura, Phys. Rev. Lett. **82**, 3540 (1999).

[4] A. Ohtomo, H.Y. Hwang, Nature **427,** 423 (2004).

[5] N. Erdman, K.R. Poeppelmeier, M. Asta, O. Warschkow, D.E. Ellis, L.D. Marks, Nature **419**, 55 (2002); T. Kubo, H. Nozoye, Surface Science **542**, 177 (2003); C. H. Lanier, Phys. Rev. B **76**, 045421 (2007); R. Herger, P.R. Willmott, O. Bunk, C.M. Schlepütz, B.D. Patterson, Phys. Rev. Lett. **98,** 076102 (2007) ; D.T. Newell, A. Harrison, F. Silly, M.R. Castell, Phys. Rev. B **75**, 205429 (2007).

[6] O.N. Tufte, Phys. Rev. **155**, 796 (1967).

[7] J.F. Schooley, Phys. Rev. Lett. **12**, 474 (1964).

[8] K. Szot, W. Speier, G. Bihlmayer, R. Waser, Nature Materials **5**, 312 (2006).

[9] T. Higuchi, T. Tsukamoto, K. Kobayashi, S. Yamaguchi, Y. Ishiwata, N. Sata, K. Hiramoto, M. Ishigame, S. Shin, Phys. Rev. B **65**, 033201 (2001).

[10] D. Kan, T. Terashima, R. Kanda, A. Masuno, K. Tanaka, S. Chu, H. Kan, A. Ishizumi, Y. Kanemitsu, Y. Shimakawa, M. Takano, Nature Materials **4**, 816 (2005).

[11] D. Kan, R. Kanda, Y. Kanemitsu, Y. Shimakawa, M. Takano, T. Terashima, A. Ishizumi, App. Phys. Lett. **88**, 191916 (2006).

[12] H. Hwang, Nature Materials **4**, 803 (2005).

[13] S. Mochizuki, F. Fujishiro, S. Minami, J. of Phys. Cond. Matt. **17**, 923 (2005).

[14] A. Rubano, D. Paparo, F. Miletto, U. Scotti di Uccio, L. Marrucci, Phys. Rev. B **76**, 125115 (2007).

[15] D. A. Crandles, B. Nicholas, C. Dreher, C. C. Homes, A. W. McConnell, B. P. Clayman, W. H. Gong, J. E. Greedan, Phys. Rev. B **59**, 12842 (1999).

[16] P. Sanchez, Materials Letters **57**, 1844 (2002).

[17] S. Kimura, Phys. Rev. B **51**, 11049 (1995).

[18] Z. Liu, Appl. Phys. Lett. **90**, 201104 (2007); J. Hiltunen, D. Seneviratne, R. Sun, M. Stolfi, H.L. Tuller, J. Lappalainen, V. Lantto, Appl. Phys. Lett. **89**, 242904 (2006).